\documentclass[fleqn,usenatbib]{mnras}

\usepackage{newtxtext,newtxmath}
\usepackage{diagbox}

\usepackage[T1]{fontenc}

\DeclareRobustCommand{\VAN}[3]{#2}
\let\VANthebibliography\thebibliography
\def\thebibliography{\DeclareRobustCommand{\VAN}[3]{##3}\VANthebibliography}

\usepackage{graphicx}	
\usepackage{amsmath}	
\usepackage[dvipsnames]{xcolor}

\newcommand{\new}[1]{{#1}}

\title[Eccentricities of BBHs with CBDs in LISA]{Eccentric Signatures of Stellar-Mass Binary Black Holes with Circumbinary Disks in LISA}

\author[I. M. Romero-Shaw et al.]{
Isobel M. Romero-Shaw$^{1,2}$\thanks{E-mail: ir346@cam.ac.uk}, Samir Goorachurn$^{3,4}$, Magdalena Siwek$^{5}$, and Christopher J.\ Moore$^{6,1,2}$
\\
$^{1}$ Department of Applied Mathematics and Theoretical Physics, Cambridge CB3 0WA, United Kingdom\\
$^{2}$ Kavli Institute for Cosmology Cambridge, Madingley Road Cambridge CB3 0HA, United Kingdom\\
$^{3}$ Physics and Astronomy Department Galileo Galilei, University of Padova, Via Marzolo, 8-35131, Padova, Italy\\
$^{4}$ Department of Physics, McGill University, 3600 University Street, Montr\'eal, QC H3A 2T8, Canada\\
$^{5}$ Center for Astrophysics, Harvard University, Cambridge, MA 02138, USA\\
$^{6}$ Institute of Astronomy, University of Cambridge, Cambridge, CB3 0HA, United Kingdom\\
}

\date{Accepted XXX. Received YYY; in original form ZZZ}

\pubyear{2023}

\begin{document}
\label{firstpage}
\pagerange{\pageref{firstpage}--\pageref{lastpage}}
\maketitle

\begin{abstract}
\new{Stellar-mass} binary black holes may have circumbinary disks if formed through common-envelope evolution or within gaseous environments.  
Disks can drive binaries into wider and more eccentric orbits, while gravitational waves harden and circularise them. 
We combine cutting-edge evolution prescriptions for disk-driven binaries with well-known equations for gravitational-wave-driven evolution, and study the evolution of stellar-mass binary black holes.
We find that binaries are driven by their disk to an equilibrium eccentricity, $0.2\lesssim e_\mathrm{eq} \lesssim0.5$, that dominates their evolution.
Once they transition to the GW-dominated regime their eccentricity decreases rapidly; we find that \new{stellar-mass} binary black holes with long-lived disks will likely be observed in LISA with detectable eccentricities $\sim 10^{-2}$ at $0.01$~Hz, with the precise value closely correlating with the binary's initial mass ratio.
This may lead \new{stellar-mass} binary black holes with CBDs observed in LISA to be confused with dynamically-formed binary black holes.
\end{abstract}




\section{Introduction}




Binary black holes (BBHs) evolving within circumbinary disks (CBDs) can be driven to long-lasting equilibrium eccentricities \citep{DOrazioDuffel2021, Zrake:2021:CBDs, Siwek:CBDs:2023}.
Merging stellar-mass BBHs form via several pathways that may involve CBDs: for example, an isolated binary may go through common envelope (CE) evolution \citep{Paczynski:1976:CE, Livio88, Bethe98, Dominik:2012:CE, Ivanova13, Kruckow16, Stevenson:2017:CE}, while a dynamically-assembled binary may evolve while embedded in an active galactic nuclei (AGN) disk \citep{Bartos:2017:AGNDisksGas, McKernan:2018:AGN, Yang:2019:AGN, McKernan:2020:AGN, Grobner:2020:AGN, Ford:2022:AGN, Samsing:2022:AGN, Calcino:2023:AGN}, and binaries evolving in field triples may also gain CBDs \citep{Silsbee16:triples, Antonini17:triples, LiuLai17:triples, Liu19:triples, VignaGomez:2021:triples, Dorozsmai:2024:triples}. 
Dynamical assembly in dense star clusters is not thought to produce BBHs with CBDs, since frequent BH partner-swappings and interactions likely disrupt disk formation \citep{Wen02:GCs, Askar:2017:GCs, Rodriguez18a, Rodriguez18b, Samsing17:GCs, AntoniniRasio:2016:NCs, Hoang:2018:NCs, Fragione19b:GCs, Chattopadhyay:2023:NCs}.
Identifying the influence of a CBD on the orbit of a BBH can therefore distinguish between different formation environments.

For BBHs detected with the current generation of ground-based detectors, measurable eccentricity is considered a robust indicator
of dynamical formation \citep{Lower:Eccentricity:2018, Zevin:2021:seleccentricity}.
Isolated BBHs are expected to have negligible eccentricities in ground-based detectors, but close-to-merger eccentricities in dynamically-formed or tertiary-driven BBH can be higher, with different pathways yielding distinct eccentricity distributions \citep[e.g.,][]{Samsing17:GCs, Antonini:2017:triples, Rodriguez18a, Tagawa:2021:AGNeccentric, ArcaSedda:2021:triples, Samsing:2022:AGN}.
\new{For example, BBHs forming in globular clusters (GCs) have $10^{-8}\lesssim e_{10}\lesssim10^{-3}$ if ejected from the cluster, $10^{-7}\lesssim e_{10}\lesssim10^{-2}$ if they merge between dynamical interactions, and $10^{-3}\lesssim e_{10}\lesssim 1$ if they merge through GW capture \citep{Zevin:2021:seleccentricity}.}  
The limited eccentricity sensitivity of Advanced LIGO-Virgo-KAGRA (LVK) \citep{AdvancedDetectorsLVK2018} detectors permits a narrow window onto eccentricity distributions: we are sensitive to only $e_{10} \gtrsim 0.05$ at GW frequencies of $10$~Hz \citep{Lower:Eccentricity:2018}.
\new{Just a small fraction of the multi-peaked distribution predicted for BBHs in GCs, which differs from those arising in other environments like AGN disks and field triples, can be used to distinguish a population of GC mergers from other channels in current detectors \citep{Romero-Shaw:2022:FourEccentric}.}

Future space-based detector LISA \citep{LISAwhitepaper} will unlock a lower-frequency stretch of the GW spectrum: $10^{-4} \lesssim f_\mathrm{GW} \lesssim 0.1$~Hz. 
LISA will be most sensitive to chirping stellar-mass BBHs approximately 10 years before they merge, at frequencies $\sim0.01$~Hz \citep{Gerosa:2019:EventRates}.
While LISA's observable volume for chirping stellar-mass BBHs is small, we do expect to capture several tens of inspiralling systems within redshifts $z\lesssim0.1$ \citep{Gerosa:2019:EventRates, Wang:2021:LISArates, Klein:2022:LISArates}.
Within the Milky Way, LISA may see $\sim5$ BBHs at lower frequencies, $\approx10^{-4}-10^{-3}$~Hz, where their slow evolution will give them quasi-monochromatic GW emission 
\citep{StrokovBerti:2024:mono}; a key challenge will be distinguishing more exotic sources like this from the dominant white dwarf binary background \citep{Moore:2023:LISAEccSensitivity}. 

Since GW frequency increases as the binary separation shrinks, LISA will observe earlier stages of a \new{stellar-mass} BBH's life \new{compared to} ground-based detectors. Since GWs circularise orbits, younger systems are also likely to have higher eccentricities \citep{Peters:1964}.
With LISA sensitivity, eccentricities $\gtrsim 10^{-3}$ can be measured for \new{stellar-mass} BBHs inspiralling through the LISA band \new{at a GW frequency of $0.01$~Hz} \citep{Klein:2022:LISArates, Garg2023, Wang:2024:ArchivalEccentric}.
Binaries evolving in isolation are expected to have $10^{-5} \lesssim e_\mathrm{0.01} \lesssim 10^{-3}$ at $0.01$~Hz, those merging after ejection from their host cluster have $10^{-4} \lesssim e_\mathrm{0.01} \lesssim 10^{-2}$, and in-cluster mergers will have $e_\mathrm{0.01}\gtrsim10^{-2}$ \citep{Wang:2024:ArchivalEccentric}.
Eccentricity is therefore likely to be a common feature of stellar-mass BBHs in LISA.

Interactions of a BBH with a CBD can cause distinct observables, which may facilitate the identification of CBD-driven BBH mergers.
Such observables include electromagnetic counterparts \citep[e.g.,][]{Mosta:2010:EMBBHCBD, Martin:2018:StellarMassBBHwithCBDs}, and mass ratios closer to unity due to preferential accretion onto the less massive component \citep[e.g.,][]{Farris2014, Gerosa:2015:PrefferentialAccretion, Duffell2020, Siwek:2023:CBDAccretionPrecession}.
As we show in this work, another feature that could distinguish \new{stellar-mass} BBHs with CBDs is distinct and measurable eccentricity values at $0.01$~Hz, $e_{0.01} \sim 10^{-2}$.
We find that the value of $e_{0.01}$ correlates with the binary's initial mass ratio $q_i$, meaning that it may also correlate with the final mass ratio $q_f$ if the BBH torques the CBD sufficiently to prevent accretion \citep{Martin:2018:StellarMassBBHwithCBDs}.
Alternatively, a population of $q_f\approx1$ BBHs may have their initial distribution of $q_i$ mapped by their detected $e_{0.01}$.

Torques from a CBD can act to contrast or amplify the effects of GW emission. 
GW emission can \textit{only} remove energy from a binary, thereby always reducing its separation and eccentricity \citep{Peters:1964}.
Analytical and early numerical simulations \citep[e.g.,][]{Pringle:1991, GouldRix:2000, Armitage2002, Haiman:2009:CBDSMBH} showed that the presence of a CBD would also facilitate inspiral. More recent hydrodynamic simulations, however, which resolve the gas dynamics \textit{within} the central cavity of the CBD, show that CBDs can cause binary inspiral \textit{or} outspiral, \citep[e.g.,][]{Miranda2016, Munoz:2019:CBDs, DOrazioDuffel2021} depending on many factors (e.g., disk thickness \citep{Tiede2020}, viscosity \citep{2020:HeathNixon:thinvsthickdisk}, binary mass ratio and eccentricity \citep{Siwek:CBDs:2023}), if the disk mass is comparable to the mass of the binary \citep{Valli:2024:CBDs}.
Hydrodynamical simulations also show that resonant disk-binary interactions typically increase the eccentricity of binaries that start close to circular \citep{GoldreichTremaine:1979:disks}. The eccentricity saturates at a limiting value \citep{LubowArtymowicz:1992:eccentricitydisk, Zrake:2021:CBDs, DOrazioDuffel2021}, the ``equilibrium eccentricity'' $e_\mathrm{eq}$, that depends on the binary mass ratio \citep{Siwek:CBDs:2023}, if the mass of the CBD is at least a few percent of the binary's mass \citep{Valli:2024:CBDs}.  

\citet{IbashiGrobner:2020:BBHinAGN} developed an analytic model to combine the driving factors of gas torques and GW emission on BBH embedded in AGN disks, with disk properties dependent on the mass and proximity of the central supermassive BH.
They estimate the eccentricities of equal-mass BBH with $M=50$~M$_\odot$ in LISA to be in the range $0.01$---$0.1$.
\citet{IshibashiGrobner:2024:AGNCBDs} updated the simplified analytic model to include the effects of accretion, studying three different total masses and five different mass ratios. 
\citet{Zrake:2021:CBDs} combine hydrodynamical simulations with equations for GW-driven evolution from \citet{Peters:1964} to evolve equal-mass BBHs \new{of masses ranging} from stellar to supermassive, finding all equal-mass BBHs have $e_\mathrm{eq}\sim0.45$\new{as a result of gas driving, and that an $M=147$~M$_\odot$ binary at redshift $z=1$ has $e \sim 5 \times 10^{-3}$ at $\sim0.01$~Hz.} \new{For other classes of LISA sources, such as the much heavier and louder supermassive BBH mergers, eccentricities can be measured at lower frequencies; for example, \citet{ArmitageKatarajan2005, Roedig2011, Siwek:2024:MBHBs} all predict that $10^{6}$~M$_\odot$ BBHs retain $e\gtrsim10^{-3}$ in LISA.}

Here, we combine cutting-edge detailed hydrodynamical simulations of BBHs with CBD \citep{Siwek:CBDs:2023} with the formulae of \citet{Peters:1964} for the evolution of orbital parameters under GW emission. We study a range of binary total masses $M \in (33, 60)$~M$_\odot$ and initial eccentricities $e_i \in (0, 0.8)$, finding that these binaries have equilibrium eccentricities between $0.2\lesssim e_\mathrm{eq} \lesssim 0.5$, depending on initial mass ratio. 
As they evolve to higher frequencies and smaller separations, GW-driven evolution takes over, and the eccentricity begins to decay rapidly.
The earlier CBD-driven evolution leaves its mark on the binary, leading to eccentricities $e_{0.01} \sim 10^{-2}$ at $0.01$~Hz.

In Sec. \ref{sec:method}, we describe our procedure for evolving stellar-mass BBH with CBDs; in Sec. \ref{sec:results}, we report our results; and in Sec. \ref{sec:discussion}, we summarise the implications of these results and discuss follow-up studies.

\section{Method}
The gas-driven evolution of BBHs is encapsulated by incremental
changes in semi-major axis, $\dot{a}_\mathrm{gas}$, and eccentricity, $\dot{e}_\mathrm{gas}$. Rates of change are derived in \citet{Siwek:CBDs:2023} for systems with different initial mass ratios, $q_i$, and initial eccentricities, $e_i$. The evolution rates are scaled by the initial separation, $a_i$, and initial total mass of the binary, $M_i$.
\new{We use the orbital evolution prescriptions published in \citet{Siwek:CBDs:2023}, computed} using a grid of hydrodynamical simulations created using the Navier Stokes version of moving-mesh code Arepo \citep{Arepo-1, Arepo-2}. The disk viscosity, $v = \alpha(h/r)^2r^2\sqrt{GM/r^3}$, is parameterized by a constant $\alpha=0.1$ \citep{ShakuraSunyaev1973}, and the disk aspect ratio is fixed at $h/r=0.1$.
The results of these simulations are in good agreement with the results of simulations using other codes that employ similar methods, e.g., \citet{Munoz:2019:CBDs, DOrazioDuffel2021, Zrake:2021:CBDs}.

The evolution rates calculated in \citet{Siwek:CBDs:2023} are valid for generic binary-plus-disk systems. Their scaling to physical quantities\new{, e.g., the absolute accretion rate and electromagnetic luminosity,} depends on the ratio of accretion efficiency in Eddington units, $f_e$, to the radiative efficiency parameter $\epsilon$. The latter may take a value between $0$ and $1$, and represents the fraction of the rest mass energy of the object lost as radiation during the accretion process. The gravitational potential energy of a particle in the disk can be lost due to friction-induced gas heating. For single BHs, $\epsilon$ ranges between about $0.06$ for Schwarzchild BHs and $0.32$ for maximally-spinning BHs \citep{LaorNetzer1989}. Interactions of BHs with each other and with their circumbinary disk, as well as poorly-constrained mechanisms of radiative feedback, make $\epsilon$ uncertain for BBH evolution. We use a conservative estimate of $\epsilon=0.1$. We note that $f_e/\epsilon=0.1$ (when $f_e=0.01, \epsilon=0.1$) could equally be achieved with, for example, $f_e=0.1, \epsilon=1$. 

Each simulated grid of BBHs comprises 90 systems with ten initial mass ratios uniformly distributed in the range $0.1 \leq q_i \leq 1.0$ and nine initial eccentricities uniformly distributed in the range $0.0 \leq e_i \leq 0.8$, matching the grid of parameters covered in \citet{Siwek:CBDs:2023}.
We evolve these binaries over a grid of ten $f_e$ values in the range $0.01 \leq f_e \leq 0.1$.
Our main results are shown for binaries of initial primary mass $m_{1,i}=30$~M$_\odot$.
Our lowest initial secondary mass is therefore $3$~M$_\odot$, a rather low BH mass that can arise through CE evolution \citep{Zevin:2020:190814} if the instability growth timescale in supernovae is assumed to be $\sim200$~ms \citep[the ``Delayed" prescription of][]{Fryer12}.
We also run otherwise-identical simulations with initial total mass $M_i=60$~M$_\odot$.

\begin{figure*}
    \centering
    \includegraphics[width=0.49\textwidth]{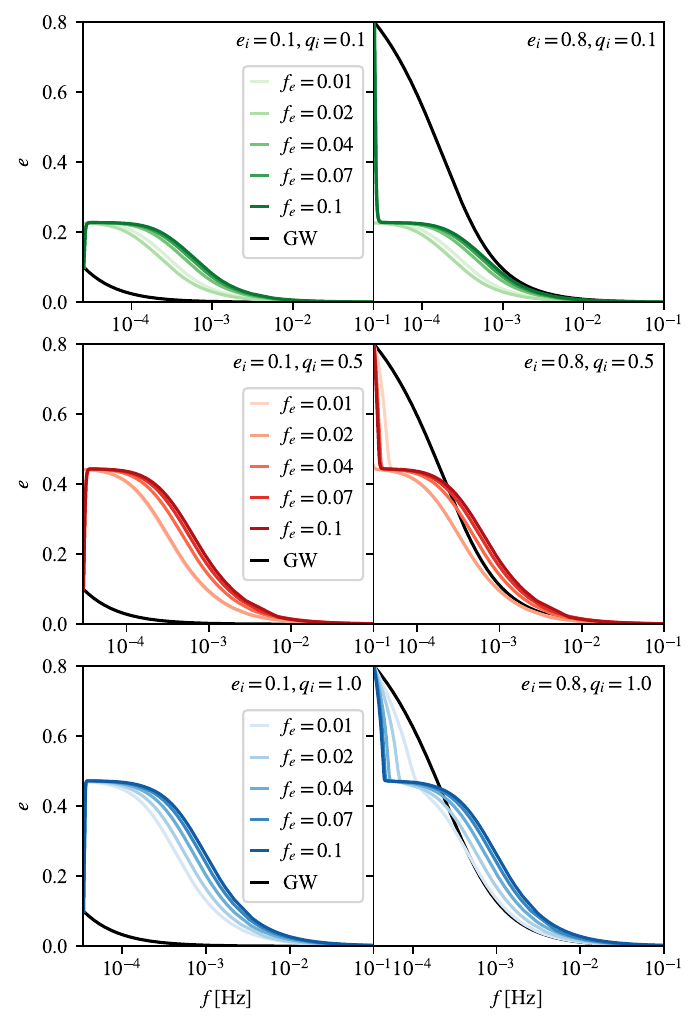}
    ~
    \includegraphics[width=0.475\textwidth]{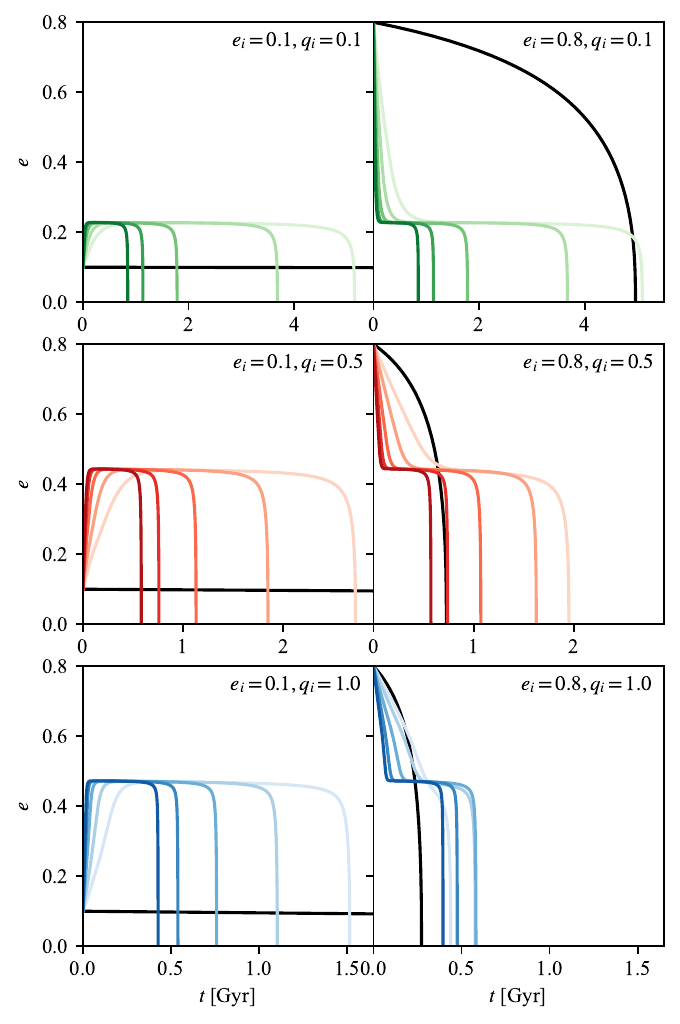}
    \caption{The eccentricity evolution of BBHs with $e_i = 0.1, 0.8$ (left and right columns, respectively, of each set of plots) and $q_i = 0.1, 0.5, 1.0$ (top to bottom rows) with initial $m_{1,i}=30$~M$_\odot$, \new{assuming \textit{unlimited} accretion (see Sec. \ref{sec:mass-accretion})}. The left set of plots shows the eccentricity evolution with peak GW frequency, while the right set of plots shows eccentricity evolution with time. The assumed $f_e$ is indicated by the opacity of the curve. 
    Because the CBD forces binaries with high $e_i$ to smaller eccentricities much more rapidly than it shrinks their separation, these binaries evolve from higher to lower peak GW frequency before reaching their steady-state.
    The steady-state eccentricity decays above $10^{-4}$~Hz.}
    \label{fig:edd_frac_varying}
\end{figure*}

We initialise each binary with a small semi-major axis of $0.2$~au to probe the regime where GWs may compete with the CBD, since GW emission only starts to influence the evolution significantly at separations on this scale.
This is also a reasonable distance to initiate BBHs immediately post-CE; indeed, separations immediately following CE can be much smaller \citep[e.g.,][]{Stevenson:2017:CE, Belczynski:2020:GW190521}.
We compare results with and without including the effects of accretion; details of the implementation are given in Section \ref{sec:mass-accretion}.

The GW-driven evolution of binary black holes is described by the equations of \citet{Peters:1964}: 

\begin{align}
    \dot{a}_\mathrm{GW} &= \frac{-64}{5} \left(\frac{G^3\mathcal{M}^3}{c^5 a^3 (1-e^2)^{\frac{7}{2}}}\right)\left(1+\frac{73}{24} e^2 + \frac{37}{96} e^4\right), \\
    \dot{e}_\mathrm{GW} &= \frac{-304}{15} e \left(\frac{G^3\mathcal{M}^3}{c^5 a^4 (1-e^2)^{\frac{5}{2}}}\right)\left(1+\frac{121}{304} e^2\right),
\end{align}
where $\dot{a}_\mathrm{GW}$ and $\dot{e}_\mathrm{GW}$ are the evolution rates of the semi-major axis $a$ and eccentricity $e$ due to GW emission, $G$ is the gravitational constant, $c$ is the speed of light, and the chirp mass $\mathcal{M} = (m_1 m_2)^{3/5} / (m_1 + m_2)^{1/5}$.

To calculate the combined effects of gas and GWs on the binary evolution, we calculate at each timestep the net change in $a$ and $e$, 

\begin{equation}
    \dot{a}_{\mathrm{net}} = \dot{a}_{\mathrm{gas}} + \dot{a}_{\mathrm{GW}}, \\
    \dot{e}_{\mathrm{net}} = \dot{e}_{\mathrm{gas}} + \dot{e}_{\mathrm{GW}}.  
\end{equation}



We estimate $f_\mathrm{GW}$ as simply double the Keplerian orbital frequency of a circular binary with separation $a$, which corresponds to the frequency of the dominant harmonic for moderately-eccentric binaries.




\label{sec:method}
\begin{figure}
    \centering
    \includegraphics[width=\columnwidth]{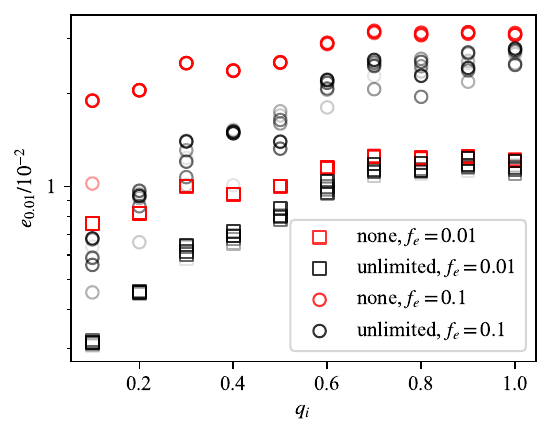}
    \caption{Eccentricities at $0.01$~Hz for all binaries with $e_i > 0.05$. Square and circular scatter points correspond to $f_e=0.01$ and $0.1$ respectively. \new{Colours correspond to the cases of \textit{unlimited acccretion} and \textit{no accretion} ("none", red markers), both with long-lived disks; binaries without disks have $e_{0.01} < 10^{-3}$}. The initial eccentricity corresponds to the alpha value of the scatter marker (higher $e_i$, higher opacity). For the negligible accretion case, $e_i$ minimally affects $e_{0.01}$, while for unlimited accretion lower $e_i$ can lead to slightly lower $e_{0.01}$. All values are detectable with LISA \citep{Klein:2022:LISArates, Wang:2024:ArchivalEccentric}.}
    \label{fig:ecc-at-001}
\end{figure}

\subsection{Accretion of gas}
\label{sec:mass-accretion}

While $f_e$ represents the fraction of the Eddington accretion rate that a single BH of mass $M=m_1+m_2$ would accrete at, the individual BHs likely accrete asymmetrically at different rates \citep{Siwek:2023:CBDAccretionPrecession}.
This parameter can be considered a proxy for the surface density of the disk, providing a measure of how much gas is available to torque the binary\new{: the denser the disk, the stronger the torque on the binary from the gas parcels that are distributed in the disk}.
The actual efficiency of mass accretion onto stellar-mass BBH from disks formed in post-CE systems is influenced by multiple uncertain factors, e.g., magnetic fields in gas disks of this type, which affect the rate of viscous transport.

\new{We consider three accretion paradigms in this work. The first is \textit{unlimited accretion}, representing a binary in an infinite gas reservoir that continually accretes mass. The second is \textit{no accretion} ("none" in Fig. \ref{fig:ecc-at-001}), representing a binary in an infinite gas disk that continually interacts with the disk but does not accrete any matter at all.  Negligible accretion may be the norm for post-CE BBHs with disks: ``typical'' binary-disk parameters for post-CE stellar-mass BBH are found to converge towards the zero-accretion case due to the torques that the binary exerts on the disk \citep{Martin:2018:StellarMassBBHwithCBDs}. The third case is \textit{limited accretion}, in which the disk mass is 10\% of $M_i$, and the binary accretes until the disk runs out.}

If the CBD contains less mass than the binary itself, accretion will deplete the disk before it has time to significantly influence its separation, although the steady-state eccentricity can be reached if the disk contains only $10\%$ of the binary's mass \citep{Valli:2024:CBDs}.
If the binary components accrete, the mass and mass ratio of the binary are time-dependent and alter the eccentricity evolution.
Mass accretion proceeds asymmetrically for unequal-mass binaries, with preferential accretion onto the lower-mass object; we implement this by interpolating the grid of $\lambda(q, e)$ values presented in Fig. 4 of \citet{Siwek:2023:CBDAccretionPrecession}. 
In our limited accretion model, the disk mass $M_d=0.1M$ and the system reverts to GW-only evolution once this is depleted. 
The influence of the disk likely decreases gradually as it loses mass to the binary and its surface density is reduced.
We approximate this behaviour by multiplying $\dot{a}_\mathrm{gas}$ and $\dot{e}_\mathrm{gas}$ by the fraction of the disk mass remaining at that timestep, $1-\delta M/M_d$.
In our unlimited-accretion model, the binary accretes at $f_e$ times its time-evolving Eddington limit throughout its evolution.

\section{Results}
\label{sec:results}

We compare examples of CBD-plus-GW-driven evolution to GW-only in Fig. \ref{fig:edd_frac_varying} for initial eccentricities $e_i=0.1, 0.8$, initial mass ratios $q_i=0.1,0.5,1.0$, and a range of $f_e$.
Consistent with \citet{Siwek:CBDs:2023}, we find that a binary is driven by its CBD towards an eccentricity that remains near-constant for a large fraction of its lifetime.
For stellar-mass BBHs with CBDs, the equilibrium eccentricity $e_\mathrm{eq}$ generally persists for $\mathcal{O}(\mathrm{Gyr})$. 

The CBD heavily influences the BBH merger timescale. 
While the GW-only merger timescale is a strong function of $e_i$, the GW-plus-gas timescale varies less with $e_i$ and more with $q_i$.
As the regulating influence of the disk is strongest when $f_e=0.1$, in this case there is almost no difference in the merger timescales of binaries with the same $q_i$ and but different $e_i$.
For low $q_i$, the merger timescales are almost identical regardless of $f_e$.

As reported in Tab. \ref{tab:LISA_ecc}, $e_\mathrm{eq}$ is set by $q_i$ for all initial eccentricities $e_i > e_{i,\mathrm{min}}$.
The values we find are consistent with those reported in 
\citet{Valli:2024:CBDs}.
These eccentricities may not be observed in BBHs with $e_i \lesssim 0.05$, which rapidly circularise in all cases except equal-mass, and those with $f_e = 0.01$, $e_i \gtrsim 0.6$ and $q_i \approx 1$, which only briefly occupy the equilibrium eccentricity before recommencing their predominantly GW-driven decay (see the lower right panel of Fig. \ref{fig:edd_frac_varying}).

\new{Stellar-mass} BBHs with CBDs are candidates for multi-band observations and archival searches in LISA data. 
Fig. \ref{fig:ecc-at-001} shows the expected eccentricity at $0.01$~Hz, $e_{0.01}$.
We find $e_{0.01}\sim10^{-2}$, within the range of detectability for such searches, and of the same magnitude as expected eccentricities from dynamically-formed mergers \citep{Wang:2024:ArchivalEccentric}.
The $e_{0.01}$ values remain set primarily by $q_i$, although for the unlimited-accretion case we expect $q_f\approx1$.
For the negligible-accretion case, in which $q_f=q_i$, this correlation a potential smoking-gun of BBHs with CBDs.
\new{We note that all $e_{0.01}$ values are given in the source frame, without considering the effects of the redshifting of the reference frequency. Using \citet{Peters:1964}, we estimate that an eccentricity of $e=3\times10^{-2}$ at $0.01$~Hz is redshifted to $e=2\times10^{-2}$ if $z=0.2$ ($10^{-3}$ if $z=1$); see also Sec. 4.1.1 of \citet{Romero-Shaw:2022:FourEccentric}. Redshifting therefore likely accounts for the much lower predicted $e_{0.01}\sim2\times10^{-3}$ for the $q_i=1$ BBH with $M_i=147$~M$_\odot$ highlighted in \citet{Zrake:2021:CBDs}, although higher-mass BBHs also circulate at lower frequencies than lower-mass BBHs.} 

The history of the binary is obscured by its interaction with the CBD. 
The worst concealment occurs for $q_i=1$, where $e_\mathrm{eq}\approx0.47$ for $e_i>0$ and $e_\mathrm{eq}\approx0.4$ for $e_i=0$; further, we find that $e_{0.01}\approx 10^{-2}$ for all $e_i$. 
Equilibrium eccentricities found in \citet{DOrazioDuffel2021, Zrake:2021:CBDs} for $q_i=1$ are also $e_\mathrm{eq}\approx 0.47$, although these studies see $q_i=1, e_i\approx0$ further circularise.
Our results are consistent with those of \citet{Siwek:CBDs:2023}, so we chalk this up to the use of different hydrodynamical simulations rather than the inclusion of GW emission.
In the final column Tab. \ref{tab:LISA_ecc}, we show $e_\mathrm{eq}$ for BBHs with $M_i=60$~M$_\odot$.
Higher-mass systems reach slightly lower steady-state eccentricities; see also \citet{Zrake:2021:CBDs, Siwek:2024:MBHBs}.

All of our simulated binaries inspiral and merge, in contrast with previous studies \citep[e.g.,][]{DOrazioDuffel2021, Siwek:CBDs:2023, Valli:2024:CBDs}, which found expanding orbits for some $e_i=0$ or low mass ratio binaries. 
This is a consequence of including GW emission: in areas of the $e_i, q_i$ plane where $\dot{a}_\mathrm{gas}$ is positive in Fig. 1 of \citet{Valli:2024:CBDs}, $\dot{a}_\mathrm{GW}$ is negative and much higher in magnitude.
Taking $e_i=0.4, q_i=0.1, f_e=0.1$ as an example, $\dot{a}_\mathrm{GW}$ is negative and $\sim10^{7}$ times higher in magnitude than $\dot{a}_\mathrm{gas}$ at $a=0.2$~au, driving the binary quickly to smaller separations.
At around $a=0.1$~au, the sign of $\dot{a}_\mathrm{gas}$ also becomes negative; however, by now the GW emission dominates even more, and $\dot{a}_\mathrm{GW} / \dot{a}_\mathrm{gas} \sim 10^{14}$.
The rapidity of eccentricity evolution for BBHs with CBDs is also explained by inspecting $\dot{e}_\mathrm{gas}$ and $\dot{e}_\mathrm{GW}$; for the same example system, both are negative, but $\dot{e}_\mathrm{gas} / \dot{e}_\mathrm{GW} \sim 6000$ at $a=0.2$~au.
This ratio becomes $\sim1$ around $a=0.1998$~au.


The GW frequency range within which a binary inhabits $e_\mathrm{eq}$ depends on $f_e$ for two reasons.
Firstly, since $f_e$ represents the availability of gas that can torque the binary, a high $f_e$ causes the binary evolution to be more strongly disk-driven and for longer.
Secondly, if we allow $\dot{m_1}, \dot{m_2} \geq 0$, the binary evolves more rapidly; in this case the mass in the disk can also be depleted rapidly, so the disk-driven epoch may be much shorter than the negligible-accretion case.
The negligible-accretion case with $f_e=0.1$ produces the highest-frequency transitions into the GW-dominated regime, leading to slightly higher eccentricities at $0.01$~Hz, as shown in Fig. \ref{fig:ecc-at-001}.

By the time the $f_\mathrm{GW}$ reaches $11$~Hz, the optimistic minimum frequency of the future ground-based detector Einstein Telescope \citep[ET;][]{Maggiore:2020:ET}, BBHs with CBDs have eccentricities 
$e_{1} \approx 10^{-4}$ if $f_e=0.1$, equivalent to $e_{10} \approx 10^{-5}$, below the expected sensitivity for ET \citep{Lower:Eccentricity:2018, Saini:2024:ETsensitivity}.

\begin{table}
    \centering
    \begin{tabular}{c||c|c|c}
    $q_i$ & $e_{i, \mathrm{min}}$ & $e_\mathrm{eq}$, $m_{1, i}=30$~M$_\odot$ & $e_\mathrm{eq}$, $M_i=60$~M$_\odot$ \\
    \hline
    \hline
    0.1 & 0.10 & 0.224 & 0.224 \\
    0.2 & 0.05 & 0.298 & 0.297---0.298 \\
    0.3 & 0.05 & 0.302 & 0.301 \\
    0.4 & 0.05 & 0.401 & 0.401 \\
    0.5 & 0.05 & 0.437 & 0.437 \\
    0.6 & 0.05 & 0.489 & 0.488---0.489 \\
    0.7 & 0.05 & 0.504 & 0.503---0.504 \\
    0.8 & 0.05 & 0.470 & 0.470 \\
    0.9 & 0.05 & 0.470 & 0.470 \\
    1.0 & 0.00 & 0.469 & 0.469 \\
    \end{tabular}
    \caption{Value of the steady-state eccentricity reached in the CDB-dominated phase of the evolution for initial eccentricities above $e_{i, \mathrm{min}}$ when the binary has $a_i=0.2$~au for varying initial mass ratios $q_i$. Below $e_{i, \mathrm{min}}$, the eccentricity reduces to $\lesssim 10^{-7}$. The values in this table are defined as the first eccentricity when $f_\mathrm{GW}\geq10^{-4}$ and  $\dot{f}_\mathrm{GW} \geq 0$. Our main results are presented for initial $m_{1,i}=30$~M$_\odot$. The $e_\mathrm{eq}$ reached for initial total mass $M_i=60$~M$_\odot$ is also shown in the rightmost column. There is slightly more variation for a higher $M_i$ because higher-mass binaries have stronger GW emission than lower-mass binaries at the same GW frequency, so begin to become GW-dominated earlier. Where there is a range, higher values of $e_\mathrm{eq}$ occur for higher $e_i$.}
    \label{tab:LISA_ecc}
\end{table}

\begin{figure}
    \centering
    \includegraphics[width=\columnwidth]{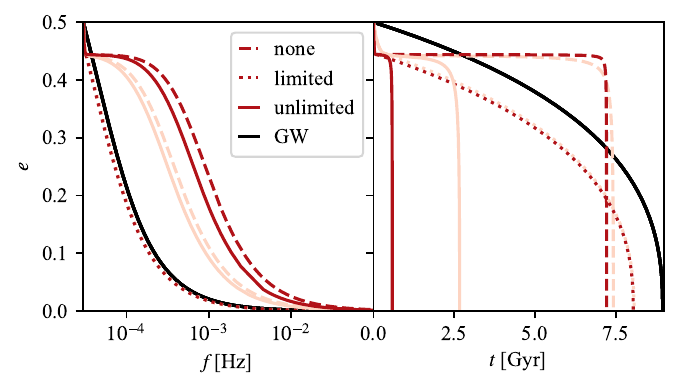}
    \caption{Frequency- (left) and time-domain (right) evolution of eccentricity for a binary with $q_i=0.5, e_i=0.5$ within different accretion paradigms. Darker and lighter colours indicate $f_e=0.1$ and $0.01$, respectively, and the accretion assumption is indicated by the linestyle. Evolution with unlimited accretion is tracked with solid curves, while results using limited accretion have dotted curves, and those assuming no accretion have dashes.
    }
    \label{fig:with-accretion}
\end{figure}

\subsection{Accretion paradigms}

Using the interpolated $\lambda(q, e)$ relationship from \citet{Siwek:2023:CBDAccretionPrecession} to split accreted mass unevenly between binary components,
there is preferential accretion onto the lower-mass object, causing the binary to evolve towards $q_f=1$.
In Fig. \ref{fig:with-accretion}, we compare the frequency- and time-domain eccentricity evolution of binaries with negligible, limited and unlimited accretion for $q_i=0.5, e_i=0.5$.
Like \citet{IshibashiGrobner:2024:AGNCBDs}, we find that unlimited accretion causes the binary to merge faster than if no accretion takes place, due to its increased mass and mass ratio.
On the other hand, because a finite disk exists for long enough to push the binary to the equilibrium eccentricity but negligibly influences the separation \citep{Valli:2024:CBDs}, limited accretion leads to a longer circularisation and merger timescale.

\subsubsection{Unlimited accretion}
\label{sec:unlimited}

In reality, this scenario requires a plentiful gas reservoir that is long-lived and allows infinite accretion, like an AGN disk. We note that (a) AGN lifetimes may be much shorter than the Gyr timescales considered here and that (b) we are not considering the additional tidal force of an SMBH. 
When accretion is unlimited, binary masses equalise and the total binary mass grows substantially over its lifetime.
Binaries finish their evolution with $q_f\approx1$ and can multiply their total mass several times; the system illustrated in Fig. \ref{fig:with-accretion} goes from $M_i=45$~M$_\odot$ to $M_f=164$ ($81$)~M$_\odot$ and $q_f=0.99$ ($0.80$) when $f_e=0.1$ ($0.01$).

As the binary continues to accrete, its mass ratio tends towards unity, pushing the turnover to GW-driven evolution to slightly lower frequencies.
Despite the mass increasing significantly during unlimited accretion, the evolution of eccentricity with frequency remains similar to the negligible-accretion case, but causes a softer transition to the GW-dominated regime.
Because binaries with higher masses and mass ratios evolve faster, the accreting binary circularises and merges several Gyr earlier than the GW-only, negligible- or limited-accretion cases.

\subsubsection{Limited accretion}
\label{sec:limited}

The initial CBD mass, $M_d$, is set to $10\%$ of the binary's total mass, an appropriate estimate for the mass contained in the CBD for post-CE systems \cite[e.g.,][]{2020:HeathNixon:thinvsthickdisk, Valli:2024:CBDs}. 
The evolution of $f_e=0.1$ and $f_e=0.01$ systems are very similar when the disk mass is limited because it is depleted within only $\sim0.04$~Gyr, and the binary then evolves under the influence of GWs only.
As shown in Fig. \ref{fig:with-accretion}, the eccentricity evolution proceeds at slightly lower frequencies than the GW-only case because the disk causes eccentricity to  be driven down faster than separation. 
This binary merges $\approx1$~Gyr faster when it has a limited CBD because of this initial rapid circularisation.
As in \citet{Valli:2024:CBDs}, a depleting disk with $M_d=0.1M$ exists for long enough to drive the BBH to the equilibrium eccentricity, but not to significantly influence its separation or mass ratio; in this example, the mass ratio grows by $0.06$ and the mass grows by $\sim4.5$~M$\odot$.

\section{Discussion}
\label{sec:discussion}

The LISA-band eccentricity evolution of a stellar-mass BBH is markedly different when it has a CBD, as the CBD drives the binary to a characteristic equilibrium eccentricity determined by its initial mass ratio. 
We find that both $e_{\rm eq}$ and $e_{0.01}$ directly correspond to $q_i$, insensitively to assumptions about accretion efficiency. 
As long as the disk endures, the eccentricity at $f_\mathrm{GW}=0.01$~Hz is $e_{0.01}\sim10^{-2}$, detectable for LISA and in the range usually expected for diskless dynamically-formed in-cluster mergers \citep{Klein:2022:LISArates, Garg2023, Wang:2024:ArchivalEccentric, DePorzio:2024:LISAEccSense}.
Since $e_{0.01}$ is a strong function of $q_i$, a population of BBHs observed with an eccentricity-mass ratio correlation would indicate contributions from non-accreting BBHs with CBDs, while a population of $q=1$ \new{stellar-mass} BBHs with eccentricities $e_{0.01}\sim2 \times (10^{-3}$---$10^{-2})$ would more likely come from accreting BBHs with CBDs. \new{The values of $e_{0.01}$ that we predict for stellar-mass BBHs with CBDs are similar to those predicted for more massive BBHs with CBDs when they enter the LISA band \citep{ArmitageKatarajan2005, Roedig2011, Zrake:2021:CBDs, Siwek:2024:MBHBs}; across all mass scales, BBHs with CBDs may be characterised by distinct mass ratios and eccentricities in LISA, once redshifting is corrected for.}

BBHs with CBDs are candidates for multi-band detections, and may be uncovered in LISA data through archival searches \citep{Wang:2024:ArchivalEccentric}.
CBDs may lead to eccentricities $\mathcal{O}(10^{-5})$ at $10$~Hz in 
ET, higher than expected from isolated binaries but below the detectable threshold for GW150914-like BBHs \citep{Lower:Eccentricity:2018}. 
The detection threshold is likely lower for lower-mass binaries \citep{Romero-Shaw:2021:TwoEccentric}, but a full population injection-recovery study is required to establish if any binaries with CBD-driven eccentricity could be identified with 3G detectors like ET. 

The eccentricity evolution of stellar-mass BBHs with CBDs is similar to that of the supermassive BBHs evolved in \citet{Siwek:2024:MBHBs}, but proceeds at higher frequencies and slightly higher eccentricities due to the lower mass. Our results using limited accretion agree with those of \citet{Valli:2024:CBDs}: disks with $M_d=0.1M_i$ do not drastically influence the separation of the BBH, but do drive the binary eccentricity to its equilibrium value.
The $e_\mathrm{eq}$ values we find for stellar-mass BBHs are in good agreement with the equilibrium eccentricity values found in \citet{Valli:2024:CBDs}, with the maximum equilibrium eccentricity of $e_\mathrm{eq} \approx 0.5$ occurring for $q_i=0.7$.
However, in contrast to previous work including \citet{Siwek:CBDs:2023, Siwek:2023:CBDAccretionPrecession, Valli:2024:CBDs}, none of our binaries undergoes orbital expansion, since $\dot{a}_\mathrm{GW} \gg \dot{a}_\mathrm{gas}$ when $a_i=0.2$~au.

CBD-driven evolution obscures the history of the binary by driving it to higher or lower eccentricity than it started with, possibly causing very different evolution pathways (e.g., CE versus dynamical formation in an AGN disk) to produce BBHs with the same characteristic eccentricities and quasi-monochromatic appearance in LISA. CBDs can also induce large changes in merger time, leading to skewed rate estimates.
However, the different expected distributions of $M_i$, $q_i$, and disk parameters for different formation channels could still lead to different population-level eccentricity distributions.
Adjusting our simulation parameters to account for these differences will be a crucial step in identifying BBHs with CBDs of alternate origin with LISA, and will be implemented in follow-up studies.  
Further work is also needed to predict $e_\mathrm{eq}$ and $e_{0.01}$ over an astrophysically-motivated $M_i$ and $q_i$ distributions for these stellar-mass BBHs, and to establish exactly how detectable these populations may be.

\section*{Acknowledgements}
The authors are grateful to the organisers of the GWPAW 2022 conference in Melbourne, Australia, at which this work was initiated. We also thank Stan DeLaurentiis, Andris Doroszmai, and Saavik Ford for useful discussions.
IMR-S acknowledges support received from the Herchel Smith Postdoctoral Fellowship Fund. CJM acknowledges the support of the UK Space Agency grant, no. ST/V002813/1.

\section*{Data Availability}
The data used to produce these results, namely the simulation outputs of \citet{Siwek:CBDs:2023}, are available upon reasonable requests made to to MS. CBD-plus-GW binary evolution codes and outputs will also be made available upon request to IMR-S.



\bibliographystyle{mnras}
\bibliography{example} 




\appendix


\bsp	
\label{lastpage}
\end{document}